\documentclass[aps,prl,twocolumn,showpacs,amsmath,amssymb]{revtex4}
\usepackage{amsfonts}
\usepackage{amsmath}
\usepackage{graphicx}
\usepackage{subfigure}
\usepackage{dcolumn}
\usepackage{bm}
\usepackage{booktabs}
\usepackage{graphicx,graphics,dcolumn,booktabs,bm}
\usepackage{longtable,lscape}
\pdfoutput=1
\usepackage{txfonts}
\usepackage{overpic}
\usepackage{amssymb}
\usepackage{indentfirst}
\usepackage{epsfig}
\usepackage{feynmf}  
\usepackage{slashed}  
\usepackage{multirow}

\usepackage{ulem}
\usepackage[usenames,dvipsnames]{color}

\usepackage[colorlinks,
            citecolor=blue,
            anchorcolor=red,
            menucolor=red,
            linkcolor=red,
            filecolor=red,
            runcolor=red,
            urlcolor=blue,
            frenchlinks=red]{hyperref}

\begin{document}

\title{Interference effect as resonance killer of newly observed charmoniumlike states $Y(4320)$ and $Y(4390)$}

\author{Dian-Yong Chen$^1$}\email{chendy@seu.edu.cn}
\author{Xiang Liu$^{2,3}$}\email{xiangliu@lzu.edu.cn}
\author{Takayuki Matsuki$^{4,5}$}\email{matsuki@tokyo-kasei.ac.jp}

\affiliation{
$^1$School of Physics, Southeast University, Nanjing 210094, China\\
$^2$School of Physical Science and Technology, Lanzhou University, Lanzhou 730000, China
\\
$^3$Research Center for Hadron and CSR Physics, Lanzhou University and Institute of Modern Physics of CAS, Lanzhou 730000, China\\
$^4$Tokyo Kasei University, 1-18-1 Kaga, Itabashi, Tokyo 173-8602, Japan\\
$^5$Theoretical Research Division, Nishina Center, RIKEN, Wako, Saitama 351-0198, Japan}

\begin{abstract}

In this letter, we decode the newly observed charmoniumlike states, $Y(4320)$ and $Y(4390)$,  by introducing interference effect between $\psi(4160)$ and $\psi(4415)$, which plays a role of resonance killer for $Y(4320)$ and $Y(4390)$.  It means that two newly reported charmoniumlike states are not genuine resonances, according to which we can naturally explain why two well-established charmonia $\psi(4160)$ and $\psi(4415)$ are missing in the cross sections of $e^+e^- \to \pi^+ \pi^- J/\psi$ and $\pi^+ \pi^- h_c$ simultaneously. To well describe the detailed data of these cross sections around $\sqrt{s}=4.2$ GeV, our study further illustrates that a charmoniumlike structure $Y(4220)$ must be introduced. As a charmonium, $Y(4220)$ should dominantly decay into its open-charm channel $e^+e^- \to D^0 \pi^+ D^{\ast-}$, which provides an extra support to $\psi(4S)$ assignment to $Y(4220)$. In fact, this interference effect introduced to explain $Y(4320)$ and $Y(4390)$ gives a typical example of non-resonant explanations to the observed $XYZ$ states, which should be paid more attention especially before identifying the observed $XYZ$ states as genuine resonances. 
  
\end{abstract}


\maketitle

{\it{Introduction}~}---~The observation of the $Y(4260)$ \cite{Aubert:2005rm} opened a new era of finding charmoniumlike states via $e^+e^-$ annihilation. Since then, a serial of $Y$ states like the $Y(4008)$ \cite{Yuan:2007sj}, $Y(4360)$ \cite{Aubert:2007zz}, $Y(4630)$ \cite{Wang:2007ea}, and $Y(4660)$ \cite{Pakhlova:2008vn} were reported in experiments, which have stimulated us to do extensive studies of their charmonium and exotic state assignments (see Refs. \cite{Liu:2013waa,Chen:2016qju} for a comprehensive review). Indeed, their experimental and theoretical progress enlarges our understanding of non-perturbative behavior of quantum chromodynamics (QCD), especially exotic hadronic matter. In the past decade, the study on charmoniumlike $XYZ$ states has become a hot issue of hadron physics. It is obvious that this wonderful story is still going on.

Very recently, two analyses \cite{Ablikim:2016qzw,BESIII:2016adj} by the BESIII Collaboration have again brought us surprise.
The precise measurements of the $e^+e^-\to \pi^+\pi^- J/\psi$ cross section at center-of-mass energies from 3.77 to 4.60 GeV \cite{Ablikim:2016qzw}
and the $e^+e^-\to \pi^+\pi^- h_c$ cross section at center-of-mass energies from 3.896 to 4.600 GeV \cite{BESIII:2016adj}
were performed by using the collected 9 fb$^{-1}$ data, where three vector structures $Y(4220)$, $Y(4320)$ and $Y(4390)$ were observed. We need to specify that the $Y(4220)$ was observed in both modes. The information of their resonance parameters in MeV is summarized below:
\begin{eqnarray*}
\begin{array}{c|ccc} \hline
{\mathrm{states}}& {\mathrm{mass}} & {\mathrm{width}}&{\mathrm{channel}}\\\hline

Y(4220)&4222.0\pm3.1\pm1.4&44.1\pm4.3\pm2.0&\pi^+\pi^-J/\psi\\

            &4218.4^{+5.5}_{-4.5}\pm0.9&66.0^{+12.3}_{-8.3}\pm0.4&\pi^+\pi^-h_c\\

Y(4320)&4320\pm10.4\pm7.0&101.4^{+25.3}_{-19.7}\pm 10.2&\pi^+\pi^-J/\psi\\

Y(4390)&4391.5^{+6.3}_{-6.8}\pm1.0&139.5^{+16.2}_{-20.6}\pm 0.6&\pi^+\pi^-h_c\\\hline

\end{array}.
\end{eqnarray*}
A new measurement of the $e^+e^-\to \pi^+\pi^- J/\psi$ cross section \cite{BESIII:2016adj} further reveals an important fact that the $Y(4260)$ contains two substructures $Y(4220)$ and $Y(4320)$.

The observation of three new vector charmoniumlike states $Y(4220)$, $Y(4320)$ and $Y(4390)$ raises some puzzles: (1) Due to the appearance of three vector charmoniumlike states at the same time, there are abundant $Y$ states accumulated at energy range from 4 to 4.5 GeV.  At present, the number of observed $Y$ states is larger than that of the predicted charmonia, which means that there is not enough space to place the observed $Y$ states in the charmonium family. Thus, we have to face how to explain the observed $Y$ states, which is a big challenge. (2) With the observation of three vector $Y$ states, we notice a puzzling phenomenon, i.e., two well established charmonia $\psi(4160)$ and $\psi(4415)$ simultaneously disappear from experimental data of cross sections for the $e^+e^-\to \pi^+\pi^- J/\psi$ \cite{Ablikim:2016qzw} and $e^+e^-\to \pi^+\pi^- h_c$ \cite{BESIII:2016adj} processes.
What are the underlying mechanisms resulting in such novel phenomena?

In this letter, we propose a unified scheme to decode three new vector charmoniumlike states $Y(4220)$, $Y(4320)$ and $Y(4390)$. By introducing the interference effect of the $\psi(4160)$ and $\psi(4415)$ in the $e^+e^-\to \pi^+\pi^- J/\psi$ and $e^+e^-\to \pi^+\pi^- h_c$ processes, the $Y(4320)$ and $Y(4390)$ structures can be well reproduced. This means that the reported $Y(4320)$ and $Y(4390)$ states can be killed. It also provides a natural solution to clarify why the $\psi(4160)$ and $\psi(4415)$ are simultaneously missing in the BESIII data \cite{Ablikim:2016qzw,BESIII:2016adj}. Under such a scheme, only the $Y(4220)$ as a genuine state is established. In the following, we address this idea in details.

\begin{figure}[htbp]
  \centering
  \includegraphics[width=8cm]{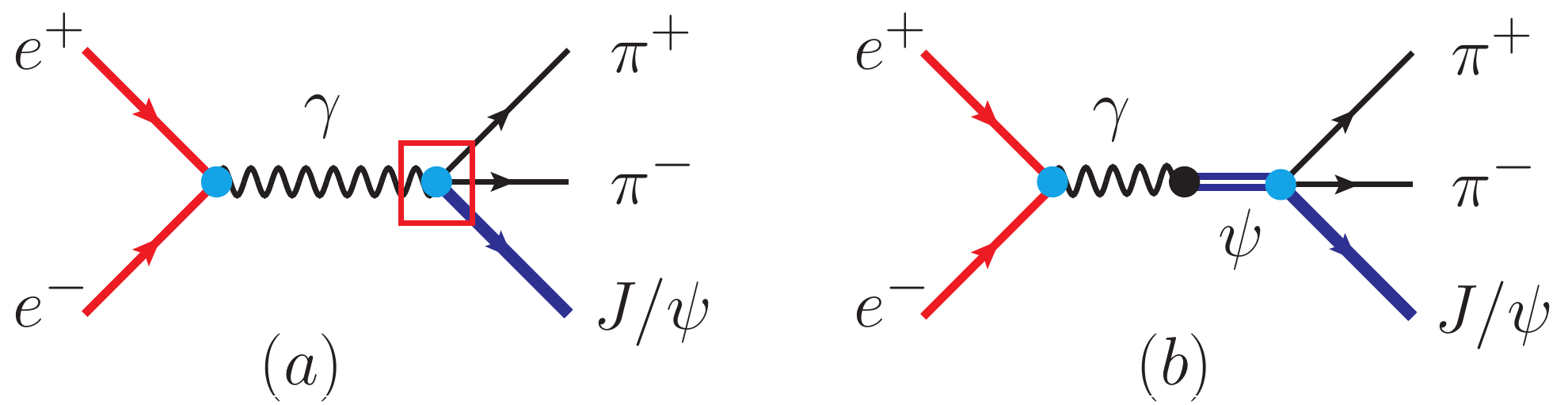}
\caption{(color online). Two schematic diagrams for the $e^+ e^- \to \pi^+ \pi^- J/\psi$ process. Here, diagrams (a) and (b) are due to nonresonance and intermediate charmonia contributions, respectively. \label{Fig:Fano}}
\end{figure}

\begin{figure}[htbp]
  \centering
  \includegraphics[width=8cm]{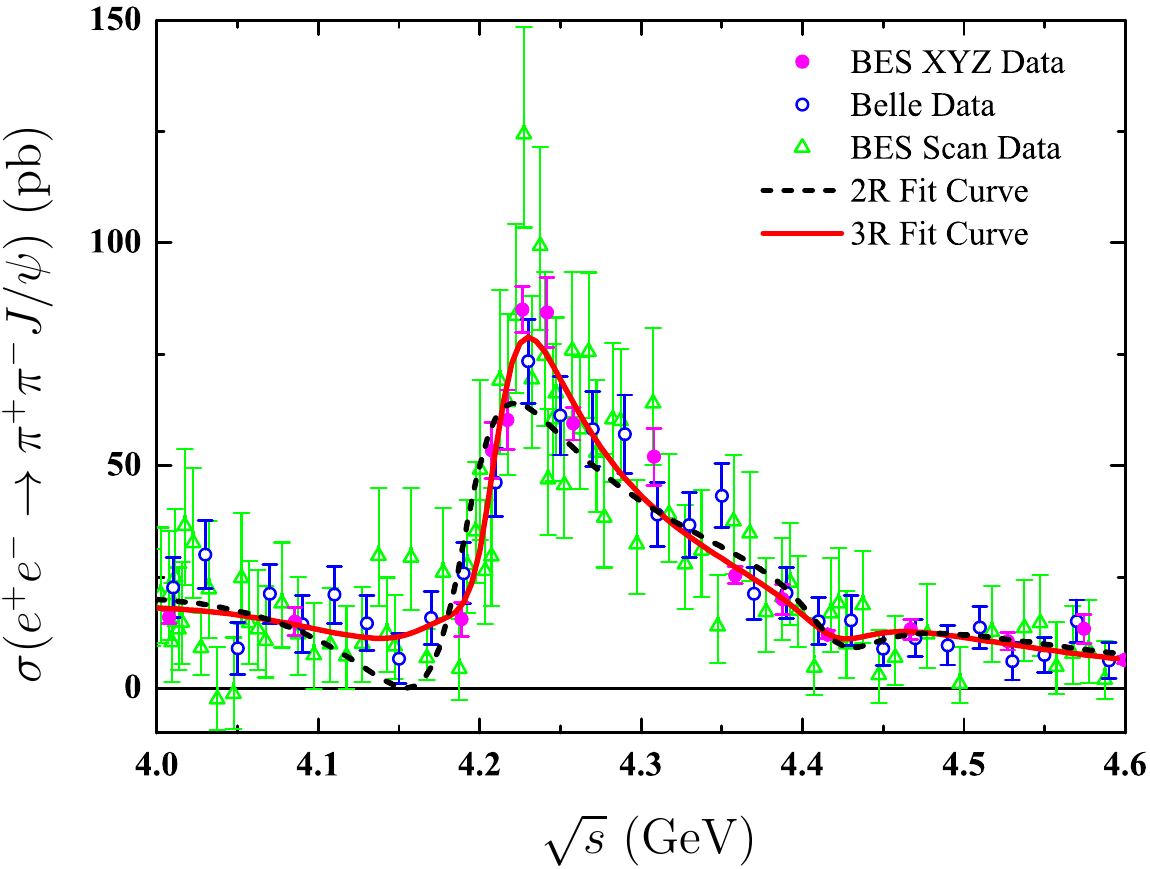}
\caption{(color online). Our fit to the cross section for the $e^+ e^- \to \pi^+ \pi^- J/\psi$ process measured by the Belle \cite{Ablikim:2016qzw} and BESIII collaborations \cite{Liu:2013dau} under the 2R and 3R fit schemes. {Here, both the BES and Belle data are included in the present fit. }\label{Fig:Jpsi}}
\end{figure}

{\it Interference effect~}---~Firstly, we focus on the $e^+ e^- \to \pi^+ \pi^- J/\psi$ process. In general, there exist two different mechanisms working in this process, which are nonresonance and intermediate charmonia contributions (see Fig. \ref{Fig:Fano}). The nonresoance contribution provides a background of the $e^+ e^- \to \pi^+ \pi^- J/\psi$ cross section, while the intermediate charmonia can reflect the corresponding signals appearing in the $\pi^+ \pi^- J/\psi$ invariant mass spectrum. However, when checking the BESIII data of the cross section for the $e^+ e^- \to \pi^+ \pi^- J/\psi$ process \cite{Ablikim:2016qzw}, we notice the appearance of a broad structure around 4.2 GeV, and the simultaneous absence of two charmonia $\psi(4160)$ and $\psi(4415)$.
What is more important is that the line shape of this broad  structure around 4.2 GeV is typically asymmetric \cite{Ablikim:2016qzw} as shown in Fig. \ref{Fig:Jpsi}, which is just sandwiched by charmonia $\psi(4160)$ and $\psi(4415)$. This peculiarity 
makes us infer the inner connection of the appearance of a broad structure around 4.2 GeV with the simultaneous absence of two charmonia $\psi(4160)$ and $\psi(4415)$. That is, we propose an interference mechanism of two charmonia $\psi(4160)$ and $\psi(4415)$ to reproduce such asymmetric broad structure existing in the $e^+ e^- \to \pi^+ \pi^- J/\psi$ process.

The amplitude of the background contribution corresponding to Fig. \ref{Fig:Fano} (a)
can be phenomenologically parameterized by
\begin{eqnarray}
\mathcal{M}_{\mathrm{NoR}} =g\, u^2 e^{-a u^2},
\label{Eq:back}
\end{eqnarray}
where $u=\sqrt{s}-\sum_{f} m_f$ is the available kinetic energy with $\sum_{f} m_f$ being the summation over the masses of all particles in the final state and $\sqrt{s}$ being the energy in the center-of-mass frame of $e^+e^-$. {The background contribution should  be a smooth curve and could be parameterized in different forms. However, different parameterization could produce very similar results  as discussed in Ref. \cite{Chen:2010nv}.  The present expression in Eq. (\ref{Eq:back}) is somehow similar to the formula for describing the background in three-body decays of B-mesons, i.e., the Argus function \cite{Albrecht:1990am}, which could reproduce the background contributions with less parameters.  The factor $e^{-a u^2}$ is introduced to balance the otherwise overestimated amplitude with increased phase space. }  In the background contributions, two phenomenological parameters $a$  and $g$ are introduced, which are obviously related to non-perturbative QCD. In the practical study, they are treated as free parameters to be determined by fitting the experimental data of the $e^+e^- \to \pi^+ \pi^- J/\psi$ cross section.
The intermediate resonance contribution corresponding to Fig. \ref{Fig:Fano} (b)
is described by a phase space corrected Breit-Wigner distribution, and the corresponding amplitude reads \cite{Ablikim:2016qzw,BESIII:2016adj},
\begin{eqnarray}
\mathcal{M}_{\mathrm{R}}(\psi)=\frac{\sqrt{12\pi \Gamma^{e^+e^-}_{\psi} \times \mathcal{B}(\psi \to \pi^+ \pi^- J/\psi)\Gamma_{\psi}}}{s-m_{\psi}^2+im_{\psi} \Gamma_{\psi} } \sqrt{\frac{\Phi_{\mathrm{2\to3}}(s)}{ \Phi_{\mathrm{2\to3}}(m_{\psi}^2)}} , \label{Eq:Res}
\end{eqnarray}
where $m_\psi$ and $\Gamma_\psi$ are the mass and width of the intermediate charmonia involved in Fig. \ref{Fig:Fano} (b). $\Phi_{2 \to 3}$ denotes the phase space of $e^+ e^- \to \pi^+ \pi^- J/\psi$. $\Gamma_{\psi}^{e^+ e^-}$ and $\mathcal{B}(\psi \to \pi^+ \pi^- J/\psi)$ are the electronic width of a resonance $\psi$ and the branching ratio of $\psi\to \pi^+ \pi^- J/\psi$, respectively. In the present study, we define $R_\psi =\Gamma^{e^+e^-}_{\psi} \times \mathcal{B}(\psi \to \pi^+ \pi^- J/\psi)$ and treat it as a free parameter. The total amplitude of the $e^+ e^- \to \pi^+ \pi^- J/\psi$ process is the sum of the nonresonance and resonance contributions, i.e., 
\begin{eqnarray}
\mathcal{M}_{\mathrm{Total}}= \mathcal{M}_{\mathrm{NoR}}+\sum_k e^{i \phi_k} \mathcal{M}_{\mathrm{R}} (\psi_k),
\end{eqnarray}
where $\phi_k$ is the phase angle between the $k$-th resonance amplitude and the nonresonace amplitude, which could be fixed by fitting the experimental data.

When introducing only two intermediate states $\psi(4160)$ and $\psi(4415)$ (2R fit scheme) interfering with a background under this scenario, we find that the line shape of the $e^+ e^- \to \pi^+ \pi^- J/\psi$ cross section can be roughly reproduced. However, 
the data around 4.2 GeV cannot be well depicted in details, which can be seen by comparing the experimental data and dashed curve corresponding to the 2R fit in Fig. \ref{Fig:Jpsi}. In Table \ref{Tab:Jpsi}, the values of fitting parameters and the corresponding $\chi^2/n.d.f$ are listed. This fact shows that a further improvement to our fit scheme is necessary. Thus, we propose 
an improved fit scheme (3R fit scheme), where one extra charmoniumlike $Y$ state is introduced associated with two charmonia $\psi(4160)$ and $\psi(4415)$ under this interference scenario. 

As shown in Fig.  \ref{Fig:Jpsi}, the BESIII data of the $e^+ e^- \to \pi^+ \pi^- J/\psi$ cross section can be well reproduced, especially for the data around 4.2 GeV (see a red fit curve in Fig.  \ref{Fig:Jpsi}).  
Under the 3R fit scheme, the obtained $\chi^2/n.d.f$ value is 118/153 smaller than that of the 2R fit scheme, which proves that the 3R fit is more suitable to depict the experimental data compared with the 2R fit scheme. The obtained fitting parameters are collected in Table \ref{Tab:Jpsi}. 
 
The above analysis indicates that the asymmetric broad structure appearing in the $\pi^+ \pi^- J/\psi$ invariant mass spectrum is due to the interference effect between two charmonia $\psi(4160)$ and $\psi(4415)$, a charmoniumlike $Y$ state and the background.  Under this scenario, we provide a reasonable solution to the puzzle of the missing $\psi(4160)$ and $\psi(4415)$ in the BESIII data of the $e^+ e^- \to \pi^+ \pi^- J/\psi$ cross section. Specifically, different form the BESIII analysis in Ref. \cite{Ablikim:2016qzw},  we only need to introduce one charmoniumlike $Y$ state, who has resonance parameters
\begin{eqnarray}
m_{Y(4220)} &=& (4207 \pm 12)\ \mathrm{MeV},\nonumber\\
\Gamma_{Y(4220)} &=& (58 \pm 38)\ \mathrm{MeV}. \nonumber
\end{eqnarray} 
In this letter, we name this state $Y(4220)$. The charmoniumlike state $Y(4220)$ 
plays a very crucial role in reproducing the details around 4.2 GeV of the line shape of the $e^+ e^- \to \pi^+ \pi^- J/\psi$ cross sections.

\begin{table}[htbp]
\caption{The parameters obtained by fitting the cross section for $e^+ e^- \to \pi^+ \pi^- J/\psi$  \cite{Liu:2013dau, Ablikim:2016qzw} and $e^+ e^- \to \pi^+ \pi^- h_c$ \cite{BESIII:2016adj}. \label{Tab:Jpsi}}
\begin{tabular}{ccc|cc}
  \toprule[1pt] \toprule[1pt]
  & \multicolumn{2}{c|}{$e^+ e^-\to \pi^+ \pi^- J/\psi$}
  & \multicolumn{2}{c}{$e^+ e^- \to \pi^+ \pi^- h_c$} \\
  Parameters
&  2R Fit
&  3R Fit
&  2R Fit
&  3R Fit \\
\midrule[1pt]
$g\ (\mathrm{GeV}^{-1})$
& $49.93 \pm  6.51$
& $49.86 \pm  5.89$
& $78.02 \pm  1.90$
& $64.84 \pm  4.75$\\
$a\ (\mathrm{GeV}^{-2})$
& $2.00  \pm  0.17$
& $2.11  \pm  0.16$
& $3.91  \pm  0.83$
& $3.41  \pm  0.21$\\
$\mathcal{R}_{\psi(4160)}\ (\mathrm{eV})$
& $5.59  \pm  0.25$
& $2.38  \pm  1.37$
& $3.62  \pm  0.29$
& $1.32  \pm  1.01$\\
$\phi_1 \ (\mathrm{rad})$
& $5.70  \pm  0.23$
& $1.59  \pm  0.76$
& $5.59  \pm  0.40$
& $5.31  \pm  0.35$\\
$\mathcal{R}_{\psi(4415)}\ (\mathrm{eV})$
& $5.14  \pm  1.82$
& $5.05  \pm  2.54$
& $1.54  \pm  0.15$
& $2.11  \pm  0.54$\\
$\phi_2\ (\mathrm{rad})$
& $4.41  \pm  0.21$
& $4.62  \pm  0.46$
& $2.93  \pm  0.62$
& $3.11  \pm  0.15$\\
$m_{Y(4220)}$
& --
& $4207 \pm 12 $
& --
& $4211 \pm 6$ \\
$\Gamma_{Y(4220)}$
& --
& $58 \pm 38$
& --
& $47 \pm 13$
\\
$R_{Y(4220)}$
& --
& $6.59 \pm 4.88$
& --
& $0.51 \pm 0.33$\\
$\phi_3$
& --
& $5.75 \pm 0.93$
& --
& $0.15 \pm 0.84$\\
\midrule[1pt]
$\chi^2/\mathrm{n.d.f}$
& 205/157
& 118/153
& 20/73
& 18/69 \\
\bottomrule[1pt]
  \bottomrule[1pt]
\end{tabular}
\end{table}

Besides getting the resonance parameters of the $Y(4220)$, we can also determine $R_{\psi}$, which is the product of the dilepton decay width and the branching ratio to $\pi^+ \pi^- J/\psi$. For $R_{\psi(4160)}$ and $R_{\psi(4415)}$, values of $R_{\psi}$ are fitted to be $(2.38 \pm 1.37)\ \mathrm{eV}$ and $(5.05 \pm 2.54)\ \mathrm{eV}$, respectively. Taking the dilepton decay width of $\psi(4160)$ and $\psi(4415)$ to be $(0.48\pm 0.22)\ \mathrm{keV}$ and $(0.58 \pm 0.07)\ \mathrm{keV}$, respectively \cite{Olive:2016xmw}, we can roughly estimate the branching ratios of $\psi(4160)\to \pi^+ \pi^- J/\psi$ and $\psi(4415)\to \pi^+ \pi^- J/\psi$ to be $(11.65 \pm 5.36)\times 10^{-3}$ and $(8.86 \pm 3.32)\times 10^{-3}$, respectively. As for the $Y(4220)$, $R_{Y(4220)}$ is fitted to be $6.59 \pm 4.88\ \mathrm{eV}$. 

\begin{figure}[htbp]
  \centering
  \includegraphics[width=8cm]{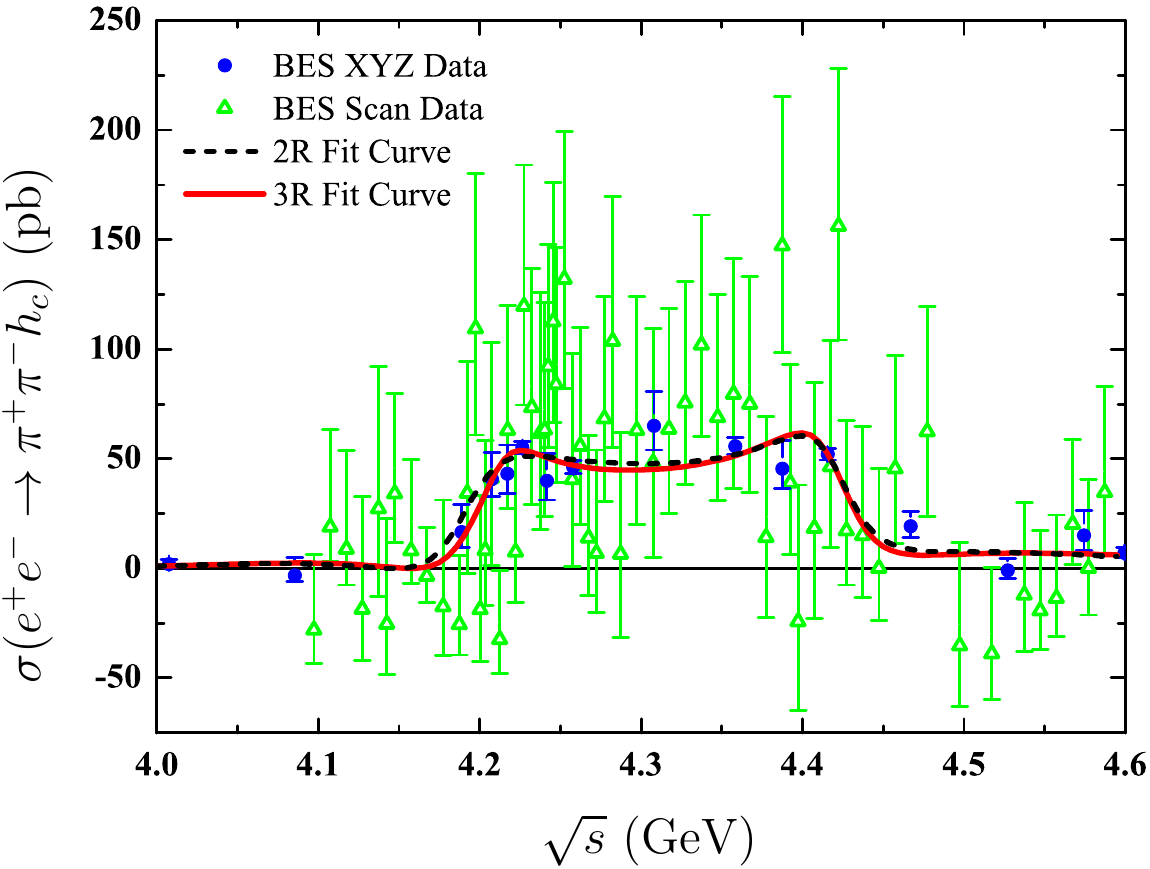}
\caption{(color online). Our fit to the cross section for the $e^+ e^- \to \pi^+ \pi^- h_c$ reported by the BES III Collaboration \cite{BESIII:2016adj}.  {Here, both the XYZ data and scan data are included in the present fit.} \label{Fig:hc}}
\end{figure}

Due to the success of introducing the interference effect to depict the $e^+e^-\to \pi^+\pi^-J/\psi$ data, we apply the same idea to study $e^+ e^- \to \pi^+ \pi^- h_c$. Here, we still firstly consider the interference among intermediate $\psi(4160)$ and $\psi(4415)$ contributions, and background. The comparison between fitting curve and experimental data is given in Fig. \ref{Fig:hc}, where the $XYZ$ data can be reproduced under this interference mechanism. 
This fact shows that introducing an extra $Y(4390)$ structure as done by BESIII seems to be redundant if considering the interference effect. Similar to the treatment of $e^+e^-\to \pi^+\pi^-J/\psi$ above, we also adopt the $3R$ fit scheme to study $e^+ e^- \to \pi^+ \pi^- h_c$ data, where the fitting parameters are collected in Table \ref{Tab:Jpsi}. Comparison of the $\chi^2/n.d.f$ values under $2R$ and $3R$ fit schemes shows that the $3R$ fit scheme cannot lead to prominent improvement when describing the experimental data as shown in Fig. \ref{Fig:hc}. Here, the resonance parameters of introduced charmoniumlike $Y$ state are 
$m=(4211 \pm 6)$ MeV and $\Gamma=(47 \pm 13)$ MeV consistent with those obtained from $e^+ e^- \to \pi^+ \pi^- J/\psi$. At present, the BESIII data of $e^+ e^- \to \pi^+ \pi^- h_c$ are not precise compared with those of $e^+e^-\to \pi^+\pi^-J/\psi$. With the accumulation of more precise data for $e^+ e^- \to \pi^+ \pi^- h_c$ in near future, we can definitely conclude whether there exists an extra charmoniumlike structure around 4.2 GeV. 
From the present $2R$ fit, we obtain $R_{\psi}$ are $(1.32 \pm 1.01)\ \mathrm{eV}$ and $(2.11\pm 054)\ \mathrm{eV}$ for $\psi(4160)$ an $\psi(4415)$, respectively. We further estimate $\mathcal{B}[\psi(4160)\to \pi^+ \pi^- h_c]=(2.75 \pm 2.45)\times 10^{-3}$, $\mathcal{B}[\psi(4415)\to \pi^+ \pi^- h_c]=(3.64 \pm 1.03)\times 10^{-3}$. Here, we notice the branching ratios for $\psi(4160)/\psi(4415)\to \pi^+ \pi^- h_c$ are smaller than those of $\psi(4160)/\psi(4415)\to \pi^+ \pi^- J/\psi$, which is consistent with the theoretical expectation of the $\psi(4160)/\psi(4415)\to \pi^+ \pi^- h_c$ decays because of spin-flip processes.

{\it{Possibility of $Y(4220)$ as $\psi(4S)$}~}---~As illustrated in this work, only $Y(4220)$ is remained as a resonance structure under interference mechanism. With this scenario, we need to further reveal the inner structure of $Y(4220)$. 

It should be noticed that a charmoniumlike structure with the mass $M=4230\pm8$ MeV and width $\Gamma=38\pm12$ MeV was discovered by BESIII in $e^+e^-\to \omega \chi_{c0}$ \cite{Ablikim:2014qwy}, which can be explained as a vector charmonium $\psi(4S)$ \cite{Chen:2014sra}. 
In Ref. \cite{Chen:2015bma}, the authors indicated that a charmoniumlike structure with mass 4243 MeV and width $16\pm31$ MeV may exist in the invariant mass spectrum of $\pi^+\pi^-\psi(2S)$ of $e^+e^-\to \pi^+\pi^-\psi(2S)$ \cite{Wang:2014hta}. A combined fit \cite{Chen:2015bma} to the data $e^+e^-\to \pi^+\pi^-\psi(2S)$ \cite{Wang:2014hta}, $h_c\pi^+\pi^-$ \cite{Ablikim:2013wzq} and $\chi_{c0}\omega$ \cite{Ablikim:2014qwy} further shows that these narrow structures around 4.2 GeV appearing in these channels can be due to the same source (a $\psi(4S)$ state), which has the mass $4234\pm5$ MeV and width $29\pm14$ MeV. Considering that the resonance parameters of $Y(4220)$ are consistent with those of the narrow structures found in channels in Refs. \cite{Wang:2014hta,Ablikim:2013wzq,Ablikim:2014qwy}, one may conclude that $Y(4220)$ studied in this work is $\psi(4S)$. 

$Y(4220)$ as $\psi(4S)$ can be supported by studies of the screen potential model \cite{Dong:1994zj, Li:2009zu} and the analysis due to the similarity between charmonium and bottomonium families,
where the $\psi(4S)$ mass is predicted to be 4273 MeV \cite{Li:2009zu} and 4274 MeV \cite{Dong:1994zj}. What is more important is that a dynamical calculation of the open-charm decay of $\psi(4S)$ by the $^3P_0$ model shows that this $\psi(4S)$ has a narrow width \cite{He:2014xna}, which provides an evidence of $Y(4220)$ as $\psi(4S)$. 

Under the $\psi(4S)$ assignment to $Y(4220)$, open-charm decay modes have a main contribution to the width of $Y(4220)$, which was illustrated by theoretical calculation in Ref. \cite{He:2014xna}. However, $Y(4220)$ has been observed only in its hidden-charm decay processes like $\pi^+\pi^-J/\psi$. 
Thus, search for a $Y(4220)$ signal in the open-charm decays becomes a crucial test if categorizing $Y(4220)$ into a charmonium family. 

\begin{figure}[htbp]
  \centering
  \includegraphics[width=8cm]{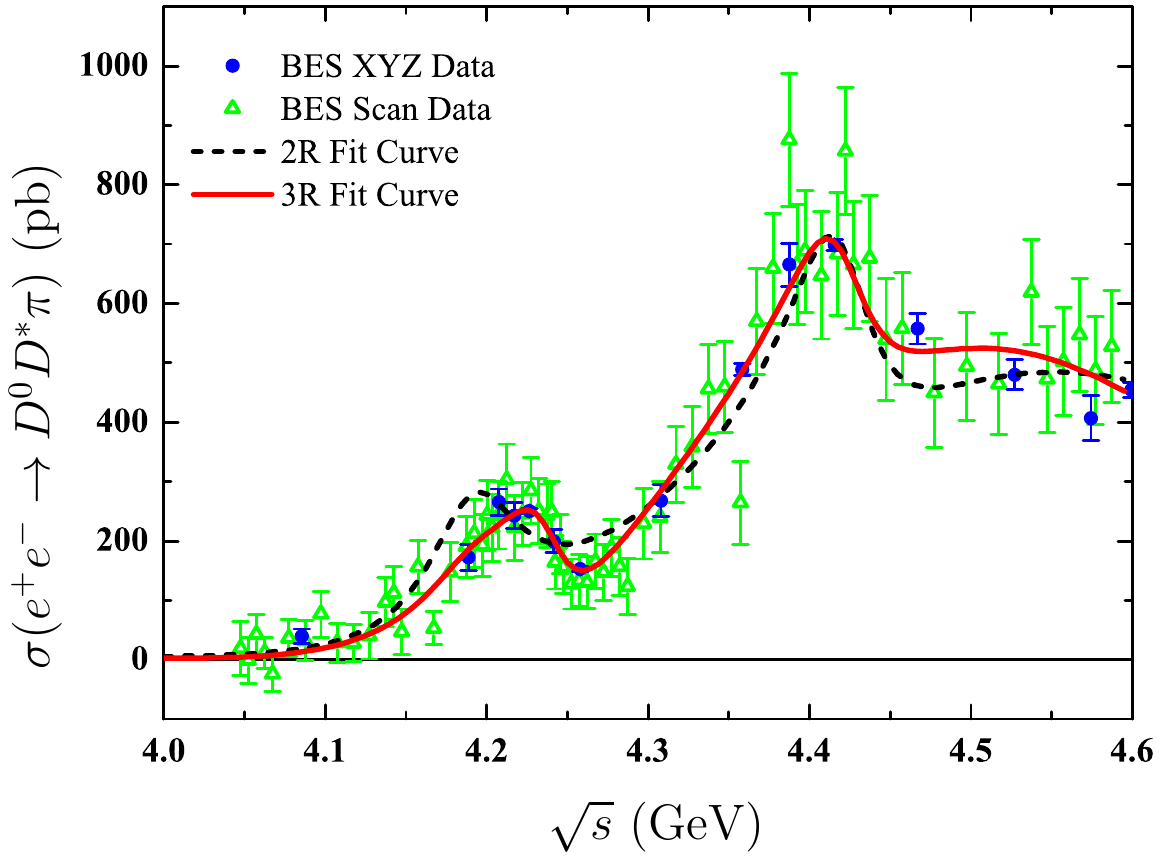}
\caption{(color online).{The same to Fig. \ref{Fig:hc} but for $e^+e^- \to D^0 \pi^+ D^{\ast -} $ process}. \label{Fig:open}}
\end{figure}

Recently, the BESIII Collaboration reported their precise measurement of the cross sections for the $e^+ e^- \to D^0 \pi^+ D^{\ast-}$ \cite{BESIII}. In the cross sections, one can find two peaks around 4.2 GeV and 4.4 GeV.  In this energy region, there are two well-established charmonium, which are $\psi(4160)$ and $\psi(4415)$. Similar to cases of the hidden charm processes, we first fit the cross sections for the $e^+ e^- \to D^0 \pi^+ D^{\ast-}$ with a nonresonance background and two resonances, which are $\psi(4160)$ and $\psi(4415)$. The fitted curves are presented in Fig. \ref{Fig:open} 
that are well fitted with experimental data. However, we have $\chi^2/\mathrm{d.o.f}=226/78$, which indicates that interference scenario of the nonresonance background and two resonances could not well reproduce the experimental data of the cross sections. By including the resonance contribution from $\psi(4S)$, the $\chi^2/\mathrm{d.o.f}$ is greatly reduced to $69/74$. From the present fit, one can find the significance of the $\psi(4S)$. The resonance parameters are fitted to be  $m=4239 \pm 26$ and $\Gamma=53 \pm 9$, which are consistent with those obtained by fitting the cross sections for $e^+ e^- \to \pi^+ \pi^- J/\psi$ and $e^+ e^- \to \pi^+ \pi^- h_c$.  In the present fit, we obtain two physical solutions of  $R_{\psi(4S)}$, which are $45.42 \pm 2.09 $ eV and $ 2.59 \pm 0.42$ eV. The former solution is about $7$ times larger than the one in the $e^+ e^- \to \pi^+ \pi^- J/\psi$, which indicates the large branching ratio of $\psi(4S) \to D^0 \pi^+ D^{\ast -}$ and is consistent with the expectation that the open charm process is the dominant decay mode of the $\psi(4S)$.

{\it{Summary}}~---~Recent observations of three charmoniumlike states $Y(4220)$, $Y(4320)$ and $Y(4390)$ announced by the BESIII Collaboration  \cite{Ablikim:2016qzw,BESIII:2016adj} brought us surprise. With the discovery of $Y(4220)$, $Y(4320)$ and $Y(4390)$, it is obvious that the family of vector charmoniumlike states is becoming more and more abundant. However, we also notice a fact that since these observed 
vector charmoniam-ike states are overcrowded in the range of 4.2-4.4 GeV, it is impossible to categorize all the observed vector charmoniumlike states into the charmonium families. It is a big challenge we have to face. 

In this work, we have proposed the interference effect to decode $Y(4320)$ and $Y(4390)$. That is, the signals of $Y(4320)$ and $Y(4390)$ existing in respective $\pi^+\pi^-J/\psi$ and $\pi^+\pi^-h_c$ invariant mass spectra can be reproduced via the interference of two well-established charmonia $\psi(4160)$ and $\psi(4415)$, and a background contribution. By this novel mechanism, we can naturally explain why the $\psi(4160)$ and $\psi(4415)$ resonance structures simultaneously disappear in the $e^+e^-\to \pi^+\pi^-J/\psi$ and $e^+e^-\to\pi^+\pi^-h_c$ cross sections. What is more important is that our study presented in this letter illustrates the interference effect plays a role of a resonance killer for newly observed charmoniumlike states $Y(4320)$ and $Y(4390)$. It means that $Y(4320)$ and $Y(4390)$ are not genuine resonances.

Fitting the experiential data under the interference mechanism mentioned above, we have also found that charmoniumlike states $Y(4220)$ must be introduced when reproducing the detailed data around $\sqrt{s}=4.2$ GeV of  the $e^+e^-\to \pi^+\pi^-J/\psi$ and $e^+e^-\to\pi^+\pi^-h_c$ cross sections. {5Here, one can notice that the cross sections for $e^+ e^- \to \pi^+ \pi^- J/\psi$, $e^+ e^- \to \pi^+ \pi^- h_c$ and $e^+ e^-\to D^0 \pi^+ D^{\ast -}$  could be, of course, well reproduced with additional resonances, $Y(4220)$, $Y(4320)$ and $Y(4390)$, as shown in Refs. \cite{Ablikim:2016qzw,BESIII:2016adj,BESIII} .  As we have discussed at the beginning of this work, there are two well established charmonia, $\psi(4160)$ and $\psi(4415)$ in the considered energy range and there is no reason to exclude these two states in the considered hidden and open charm processes.  Thus, we have firstly attempted to reproduce  the experimental data with a non-resonance background and two charmonia, i.e., $\psi(4160)$ and $\psi(4415)$. We have found if only these two charmonia are included, the experimental results can not be well described.  As the next attempt, we have further included the charmonium-like state $Y(4220)$ since this state was observed in more than one channel, such as  $e^+ e^- \to  \chi_{c0} \omega$ and $e^+ e^- \to \pi^+ \pi^- \psi(2S)$. With a non-resonance background and three resonances, we have found that the experimental data for both hidden-charm and open charm processes can be well reproduced in the same scenario. From the present fit, we can conclude that the experimental data can be reproduced with two well established charmonia $\psi(4160)$ and $\psi(4415)$ and one charmonium-like states $Y(4220)$, while the charmonium-like states $Y(4320)$ and $Y(4390)$ are not necessary.   }

In this letter, we have also discussed the possibility of the observed charmoniumlike state $Y(4220)$ as a charmonium $\psi(4S)$. As a charmonium, $Y(4220)$ should dominantly decay into open-charm channels. We have noticed the experimental result of the $e^+ e^- \to D^0 \pi^+ D^{\ast-}$ cross section \cite{BESIII}. Further fitting this experimental data, we have indicated the evidence of a charmoniumlike state $Y(4220)$ in $D^0 \pi^+ D^{\ast-}$ invariant mass spectrum, which provides an extra support to $Y(4220)$ as a charmonium $\psi(4S)$.  With more precise experiment data collected by future experiment, this charmonium assignment to $Y(4220)$ can be tested. 

In summary, in the past decade, theorists have paid more attentions to resonance explanation to such abundant experimental observations of charmoium-like states, which were assigned to different charmonium or exotic state. 
In this letter, we have explicitly emphasized that the non-resonance explanations to the observed charmoium-like states should not be disregarded before establishing the reported charmoniumlike states as genuine resonances. The interference effect introduced in this letter can be a very typical example of non-resonance explanations to the observed charmoium-like states. In the future study of $XYZ$ charmoniumlike states, we also encourage our colleagues to focus on non-resonance mechanism, which may provide a unique perspective beyond resonance explanations to $XYZ$ states.

\section*{ACKNOWLEDGMENTS}

We would like to thank Professor Chang-Zheng Yuan for useful discussion. This project is supported by
the National Natural Science Foundation of China under Grants Nos. 11222547, 11175073, 11775050 and 11375240 and
the Ministry of Education of China (the Fundamental Research Funds for the Central Universities). Xiang Liu is also supported by the National Program for Top-notch Young Professionals.

\end{document}